\documentclass[twocolumn,epjc3]{svjour3}

\RequirePackage{amsmath}
\RequirePackage{amssymb}
\RequirePackage{flushend}
\RequirePackage[numbers,sort&compress]{natbib}

\journalname{Eur. Phys. J. C}

\newcommand{\be}{\begin{equation}}
\newcommand{\ee}{\end{equation}}

\newcommand{\no}{\nonumber\\}
\newcommand{\ba}{\begin{eqnarray}}
\newcommand{\ea}{\end{eqnarray}}
\newcommand{\ci}[1]{\cite{#1}}
\newcommand{\bi}[1]{\bibitem{#1}}

\begin{document}

\title
{ Gravity effects on thick brane formation from scalar field dynamics}
\author{Alexander A. Andrianov\thanksref{e1,addr1,addr2}
    \and
    Vladimir A. Andrianov\thanksref{e2,addr1}
    \and
    Oleg O. Novikov\thanksref{e3,addr1}
}

\thankstext{e1}{e-mail: andrianov@icc.ub.edu}
\thankstext{e2}{e-mail: v.andriano@rambler.ru}
\thankstext{e3}{e-mail: oonovikov@gmail.com}

\institute{V.A. Fock Department of Theoretical Physics, Saint-Petersburg State University, ul. Ulianovskaya, 198504 St. Petersburg, Russia\label{addr1}
\and
Institut de Ci\`encies del Cosmos, Universitat de Barcelona, Mart\'i Franqu\`es, 1, E08028
Barcelona, Spain\label{addr2}
}

\date{Received: date / Accepted: date}

\maketitle

\begin{abstract}
The formation of a thick brane in
five-dimen\-sional space-time  is investigated when warp geometries of
$AdS_5$ type are induced by scalar matter dynamics and triggered by a thin-brane defect.  The scalar matter is taken to consist of two fields with $O(2)$ symmetric self interaction and with manifest $O(2)$ symmetry breaking by terms quadratic in fields. One of them serves as a thick brane formation mode around a kink background and another one is of a Higgs-field type which may develop a classical background as well. Scalar matter interacts with gravity in the minimal form and gravity effects on (quasi)localized scalar fluctuations are calculated  with usage of gauge invariant variables suitable for perturbation expansion.  The calculations are performed in the vicinity of the critical point of spontaneous breaking of the combined parity symmetry where a non-trivial v.e.v. of the Higgs-type scalar field is generated. The nonperturbative discontinuous gravitational effects  in the mass spectrum of light localized scalar states are studied in the presence of a thin-brane defect. The thin brane with negative tension happens to be the most curious case when the singular barriers form a potential well with two infinitely tall walls and the discrete spectrum of localized states arises completely isolated from the bulk.
\end{abstract}


\section{\label{sec:intro}Introduction}
The embedding of our universe in  four-dimensional space-time into a higher-dimensional space-time has attracted recently much interest \ci{rushap},\ci{otherbr}-\ci{otherbr3} as a plausible realization explaining the weakness of gravity \ci{ADD} and providing a way \ci{RSI} \ci{RSII} to resolve the hierarchy problem. Different
applications of such a construction to particle physics, astrophysics and cosmology can be
found in reviews \ci{RuBar} - \ci{loc6}.

In the minimal approach the matter universe is supposed to exist in five-dimensional space-time with localization  of relatively light matter fields in the vicinity of three-dimensional hypersurfaces (thick 3-branes). The important problem is what could be a localization mechanism and whether it can be provided by nonlinear interaction of a multidimensional matter resulting in spontaneous translational symmetry breaking which could create domain walls \ci{RuBar},\ci{RuBar1}, \ci{rev121} -\ci{rev14}.
The dynamics of matter localization on domain walls happens to be essentially influenced  by gravity \ci{rev121} -\ci{rev19}.

In this work we continue our brief investigation \cite{partnuc} of  the formation of domain walls by two scalar fields with nonlinear interaction. These fields minimally couple to
gravity in five-dimensional space-time \ci{aags2} and induce  asymptotic Anti-de Sitter geometries far from the brane.
If gravity effects are neglected then localization of light
particles on a thick brane can be well realized with the help of a
background scalar  configurations with nontrivial topology ("kinks")  \ci{aags1} . However  it was argued in \ci{rev15}, \ci{rev18}, \ci{rev19} that for one scalar field, gravity induces singular repulsion towards the remote AdS horizon so that
localized modes on a brane may be absent and a massless Goldstone-type mode of translational symmetry breaking disappears.

One of the purposes of this work is to search for brane localized states when the matter sector includes two scalar fields with $O(2)$ symmetric self-interaction which are mixed with gravity scalar modes. Eventually one finds that among scalar mode fluctuations one is a "branon" \cite{branon}) eventually related to brane deformations  and another one is responsible for  fermion mass generation (similar to a Higgs field) \ci{aags2}. Spontaneous breaking of translational symmetry  is provided by a Higgs-type mechanism with terms in the potential softly breaking $O(2)$ symmetry ("tachyon masses" of scalar fields). In the tachyon mass parameter space a critical point of spontaneous parity breaking appears which is important to obtain the v.e.v. of the Higgs-like scalar field. The latter plays an indispensable role to give masses for fermions  localized on domain wall \ci{aags2}, \ci{loc1} -\ci{loc5}.

Thereby the two phases arise: in the first phase the only nontrivial v.e.v. is given by a kink-like scalar field configuration. However the branon fluctuations around such a kink in the presence of gravity are suppressed by the repulsive singular potential  which appears to violate gravitational perturbation theory \ci{rev19}.  Meantime the Higgs-like field in this phase decouples from branons, becomes massive and exhibits a more regular weak gravity behavior.
In the second phase the Higgs-type field obtains a nontrivial v.e.v. suitable for generation of fermion masses \ci{aags1}. Both fields, branons and Higgs-like scalars as well as the scalar gravity mode are mixed. In this phase the  Higgs-like mass exhibits a discontinuity in the limit of vanishing gravity as compared with the two scalar field model without gravity.

In this work we switch on a four-dimensional space-time defect in the form of a rigid thin brane with small tension. It breaks manifestly the invariance under five-dimensional diffeomorphisms, in particular, translational invariance along the fifth dimension, and it is assumed to trigger spontaneous translational symmetry breaking (kink formation) around a chosen hyperplane.

In Sect. \ref{sec:model} we remind the model of two scalar fields  with their minimal
coupling to gravity  for arbitrary potential and derive the equations of motion. This model is supplemented with a thin-brane defect characterized by a tension.
Its presence leads to the necessity of inclusion of a so-called Gibbons-Hawking-York term on the thin brane hyperplane.
In subsect. \ref{sec:model_minimal} the scalar potential is restricted to a quartic $O(2)$ symmetric potential
and supplemented by soft breaking of  $O(2)$ symmetry quadratic in fields ( as
it could arise from the fermion induced effective action \ci{aags2}). For this
Lagrangian the gaussian normal coordinates are introduced and the appropriate
equations of motion are obtained. The existence of two phases  which differ in
presence or absence of v.e.v. for the Higgs-like field  is revealed and the
solutions for classical background of both scalar fields are found  in the
leading approximation of the gravity coupling expansion. In  Subsection 2.3
the next-to-leading approximation is performed in the phases with unbroken and
broken combined parity symmetry correspondingly.

In Sect.\ref{sec:fluct} the action of scalar fields and
gravity  is calculated in a vicinity of a background metric up to quadratic order in fluctuations. It is found
 in gauge invariant variables suitable for perturbative expansion both in gravitational constant and in thin-brane tension.

 In Sect.\ref{sec:phichannel} the mass spectrum of the
 branon fluctuations is thoroughly investigated in the phase with unbroken
 combined parity symmetry for different values of the thin brane tension.
 Because the emergence of the repulsive singular barrier is related
with vanishing first derivative of the metric factor  the presence of a space-time defect in the form of a thin brane with positive tension regularizes a gravity born singular barrier and it really makes possible for localized massless  and  heavy modes
to appear. Still the discontinuity in the zero-gravity limit does not go away completely and has to be accounted for.
 In Subsect. \ref{sec:phichannel_negbrane} thin branes with negative tension are considered. It is the most curious case as the singular barriers form a potential well with two infinitely tall walls. Thus we obtain the infinite discrete spectrum of localized states completely isolated from the bulk.

In Sect.\ref{sec:lightscalar} we use perturbation theory around the critical point in the weak
gravity expansion to obtain the mass of  Higgs-like boson in the presence
of nonperturbative effects  and reveal another discontinuity in the mass spectrum.
In concluding section, we discuss results and
prospects of the proposed model.

\section{\label{sec:model}Formulation of the model in bosonic sector}
\subsection{\label{sec:model_general}General two-boson potentials: gaussian normal coordinates}
We want to examine  the properties of scalar matter localization on a 3-brane in five-dimensional  space with taking  its own gravity into account.  As it was found in \cite{rev15,rev19} the spontaneous localization of scalar matter on a 3-brane is governed by a gravity induced singular barrier. However the  spontaneous brane generation physically is not fully consistent. One has to justify the brane arising centered in a given point in the fifth direction, i.e. a (presumably) small defect in space geometry must be introduced to trigger a brane creation in the specified point.

To perform the above program let us endow the five-dimensional space with gravity describing it by a pseudo Riemann metric tensor
 $g_{AB}$. In flat space and for the rectangular coordinate system this tensor is reduced  to the Minkowski metric $\eta_{AB}$. The extra-dimension coordinate is taken space-like,
\ba
&&(X_\alpha) = (x_\mu, y)\ , \quad (x_\mu) = (x_0, x_1, x_2, x_3),\\
&&(\eta_{\alpha\alpha}) = (+,-,-,-,-)
\ea
and the subspace of coordinates $x_\mu$ corresponds to the
four-dimen\-sional Minkowski
space.

We confine ourselves to the dynamics of two real scalar fields $ \Phi(X)$ and $ H(X)$ with  minimal interaction to gravity and supplement it with the geometry defect of thin brane type  stuck to the point $y=0$ along the extra dimension. The corresponding action functional \cite{partnuc} reads,
    \ba
S[g, \Phi,  H ]&=& \int {d^5 X}\sqrt {\left| g \right|} \left(- \frac12 M_\ast ^3  R+{\cal L}_{mat}(g, \Phi,  H )\right) \no &&- 3 M_\ast^3 \lambda_b \int\limits_{y=0} d^4X \sqrt{\left| ^{(4)}g\right|} +S_{GH}, \label{1} \ea
    \be
{\cal L}_{mat} = Z\left(\frac12(\partial _A  \Phi \partial ^A  \Phi +\partial _A  H \partial ^A  H)  - V\left(  \Phi,  H  \right)\right) , \ee
where $R$ stands for a scalar curvature, $ \left| g \right| $ and $\left| ^{(4)}g\right|$ are the determinants of the 5-dim and induced 4-dim metric tensors correspondingly the latter being determined  in the gaussian coordinate system. $ M_\ast $ denotes a five-dimensional gravitational Planck scale. The thin brane defect is taken in the form of cosmological  constant on a {\it rigid} 3-brane parameterized by $ \lambda_b$. Thus for this type of defects the thin brane fluctuations are suppressed. As it has been realized in \cite{GW,york} the consistent dynamics of gravity and matter in the presence of a brane needs in adding into the thin brane action a compensating Gibbons-Hawking-York term which is included into eq. \eqref{1} as $S_{GH}$. It is expressed in terms of the affine connection $\Gamma_{BD}^C$,
\be
S_{GH}=\frac{1}{2}M_\ast^3 \int\limits_{y=0} d^4X \sqrt{\left| ^{(4)}g\right|} \left[\Gamma_{\mu 5}^\mu g^{55}-\Gamma_{\mu\nu}^5 g^{\mu\nu}\right]_\pm
\ee

In the scalar matter action the normalization coefficient $Z$ has dimension of mass and is introduced to simplify the  equations of motion.

The related Eqs. of motion in the bulk aside the thin brane, $y\not= 0$, take the following form,
    \ba&&
R_{AB}  - \frac{1}{2}g_{AB} R = \frac{1}{{M_\ast^3 }}T_{AB} ,\\    &&  D^2  \Phi  =  - \frac{{\partial V}}{{\partial  \Phi }} ,\quad  D^2  H  =  - \frac{{\partial V}}{{\partial  H }},
 \ea where $D^2$ is a covariant D'Alambertian, and the energy-momentum tensor reads,
    \ba
&&T_{AB}  = Z \Bigl\{\partial _A  \Phi \partial _B  \Phi + \partial _A  H \partial _B  H-\no
&&- g_{AB} \left(\frac12 \Big(\partial _C  \Phi \partial ^C  \Phi + \partial _C  H \partial ^C  H\Big)  - V\left(  \Phi,  H  \right) \right)\Bigr\}. \ea
In order to build a thick $3 + 1$-dimensional brane we  search for classical vacuum configurations which do not violate 4-dimensional Poincare invariance. In this section a background solution for the metric is searched for in the gaussian frame, $$ds^2 = e^{-2\rho(y)} \eta_{\mu\nu}dx^\mu dx^\nu - dy^2. $$ This kind of background  metric suits well for interpretation of scalar fluctuation spectrum and corresponding resonance effects (i.e. scattering states) \ci{aags2}.

For such a metric aside the thin brane, $y\not= 0$, the equations of motion become,
 \ba &&\rho''= \frac{Z}{3M_\ast^3} (\Phi'^2 +  H'^2),\label{eoms0}\\
 &&\frac{2Z}{3M_\ast^3} V( \Phi,  H) =  \rho''-4(\rho')^2, \\
 &&\Phi''-4\rho'\Phi'= \frac{\partial V}{\partial \Phi} ,
\quad H''-4\rho'H'= \frac{\partial V}{\partial  H}. \label{eoms}\ea
One can prove \ci{aags2}, that only three of these equations are independent.

These bulk equations should be supplemented with the Israel matching conditions on the thin brane that for our ansatz read,
\ba
&&\left[\rho'\right]_\pm=2\lambda_b,\quad \left[\Phi'\right]_\pm=0,\\
&&\left[\Phi''\right]_\pm = 8\lambda_b \Phi'(0),\quad \left[H'\right]_\pm=0.
\ea
Their consistency is provided by the Gibbons-Hawking-York compensating term in the action.

We will consider the potentials analytic in scalar fields and bounded from below which exhibit the discrete symmetry under reflections $ \Phi \longrightarrow  -\, \Phi$ and $ H \longrightarrow -\, H$ and have few  minima  for non-vanishing v.e.v. of scalar fields. Thus there exist constant background solutions $\{ \Phi_{min},\  H_{min}\}$ which are compatible  with the Einstein equations provided that the vacuum energy
\ba Z\langle V ( \Phi,  H)\rangle &=& Z V(\{ \Phi_{min},\  H_{min}\})\no &\equiv&  -\Lambda_{c} M^3_\ast > 0,\ea i.e. for the negative cosmological constant $\Lambda_{c}$. In this case a classical geometry will be of Anti-de-Sitter type.
\subsection{\label{sec:model_minimal}Minimal realization in $\phi^4$ theory}
 Now let us restrict ourselves with studying the formation of a brane in the theory with a minimal potential bounded from below and admitting kink solutions which connect two potential minima. The potential consists of a quartic scalar self-interaction
and (wrong-sign) mass-like terms for both scalar fields.  This potential is constructed to possess the  $O(2)$-symmetry of vertices quartic in fields but quadratic couplings are taken manifestly breaking $O(2)$-symmetry. The effective action of scalar matter looks as follows,
\ba
{\cal L}_{mat}&=&\frac{3\kappa}{2M^2}\Big(\partial _A  \Phi \partial ^A  \Phi +\partial _A   H \partial ^A   H \no
&&+2M^2  \Phi^2 +2\Delta_H   H^2- ( \Phi^2 +  H^2)^2 -  V_0\Big), \label{36}
\ea
where the normalization of the scalar field lagrangian is chosen as $Z = 3\kappa M_\ast^3/M^2$ for simplification of Eqs. of motion and of the gravitational perturbation expansion.
To associate it to the weak gravity limit we specify that  $ \kappa \sim M^3/M_\ast^3 $ is a small parameter $ \kappa\ll 1$ which
characterizes the interaction of gravity and matter fields. Let us fix $M^2 > \Delta_H $ then the absolute minima correspond to $ \Phi_{min} = \pm M,\   H_{min} = 0 $ and a constant shift of the potential energy must be set $V_0 = M^4$ in order to determine properly the 5-dim cosmological constant $\Lambda_{c}$.

The classical equations \eqref{eoms} for this model contain the terms of different orders in small
parameter  $ \kappa $, and  they can be solved in
perturbation theory assuming that,
\be
\frac{{\left| {\rho '(y)} \right|}}{M} = O(\kappa ) = \frac{{\left|
{\rho ''(y)} \right|}}{{M^2 }},\quad \frac{\lambda_b}{M}=O(\kappa) .
\ee
Thus in the leading order in $ \kappa $ the equations for the fields $
 \Phi \left(y \right) ,   H  \left(y \right) $ do not depend on the metric factor $\rho(y)$ and  on the thin brane tension $\lambda_b\ll M $.

Depending on the relation between quadratic couplings $M^2$ and $\Delta_H$ there are two types of solutions of
eqs. \eqref{eoms} inhomogeneous in $y$  \cite{aags1}. In the zero gravity and thin-brane tension limit the first solution dominates when $\Delta_H \leq M^2/2$,
    \be
 \Phi  = \pm M\tanh \left( {My} \right) + O\left( {\kappa } M, \lambda_b \right),\quad H(y) = 0 .  \label{kink0}\ee
To the leading order in $\kappa\sim \lambda_b$ it generates the following conformal factor ,
    \ba
\rho_1 \left( y \right) &=& \frac{2{\kappa }}{3}\left\{ {\ln \cosh
\left( {M y} \right)+\frac{1}{4}  \tanh^2( M y)}\right\}\no
&&+\lambda_b |y|+ O\left(
{\kappa ^2, \kappa\lambda_b/M, \lambda_b^2/M^2 } \right) , \label{rho0}\ea
which is chosen to be an even function of $y$ in order to preserve the so-called $\tau$ symmetry which has been mentioned above as a combined parity symmetry. The spontaneous breaking of this symmetry is associated with fermion mass generation \ci{aags1,partnuc}. The very symmetry involves the intrinsic parity reflection of both fields and the reflection of fifth coordinate,
$$\Phi(y) \to -\Phi(- y),\quad H(y) \to - H(-y).$$ Evidently the parity reflection leaves the bo\-son\-ic action \eqref{36} invariant and holds as a symmetry for the kink \eqref{kink0}. In the presence of gravity induced by a background matter the $\tau$ symmetry survives for even conformal factors.

For the spectrum of metric fluctuations the essential role is played by the conformal factor derivative \cite{partnuc}. Namely zeroes of $\rho'(y)$ generate singular barriers $\sim 1/(\rho'(y))^2$ in the mass operator of scalar-mode fluctuations. Let us analyze different options with the profile \eqref{rho0}. Without a defect $\lambda_b = 0$ evidently $\rho'(0) = 0$ and the singular barrier arises at the origin \cite{partnuc}. If the constant $\lambda_b$ is positive then the function $|\rho'(y)|$ does not possess any zeroes (apart from a marginal one at $y = 0$ of zero measure), correspondingly $|\rho'(0)| = \lambda_b$. The absence of singular barrier opens the possibility to localize a scalar zero-mode as it will be shown in Sect. \ref{sec:phichannel}.

On the other hand one can consider a negative brane tension $\lambda_b < 0$. In order to retain an asymptotic AdS geometry one has to provide the positive AdS curvature $k = {\frac23 \kappa M + \lambda_b >0}$, i.e. $|\lambda_b|/M < \frac23 \kappa$. Meantime the function,
\be\rho_1'(y) = \kappa M  \left(\tanh My - \frac13 \tanh^3 My\right) - |\lambda_b| \epsilon(y), \ee
where $\epsilon(y)$ is a sign function, possesses two zeroes symmetric in respect to $y=0$. These zeroes create two singular barriers splitting the fifth dimension into three regions with nearly independent mass spectra. In the region between the barriers one can reveal an infinite tower of localized scalar states with discrete masses as it will be derived in the Subsect. \ref{sec:phichannel_negbrane}.

The second kink profile arises only when $M^2/2 \leq \Delta_H \leq M^2$ (in the zero gravity and thin-brane tension limit), i.e. $2\Delta_H = M^2 +\mu^2, \ \mu^2 < M^2$,
\ba
&&\Phi_0(y)  = \pm M\tanh \left( {\beta M y} \right), \no
&& H_0 (y) = \pm \frac{\mu}{\cosh \left( {\beta M y} \right)} ,\no
&&\beta = \sqrt{1 - \frac{\mu^2}{M^2}}, \label{zeroap} \ea
and it breaks the $\tau$ symmetry. Therefrom one can find the conformal factor to the leading order in $\kappa, \lambda_b$ in the following form,
\[
\rho_1 \left( y \right) = \frac{{\kappa }}{3}\left\{ \left(3 -  \beta^2\right) {\ln \cosh
\left( {\beta M y} \right)+\frac{1}{2} \beta^2 \tanh^2(\beta M y)}\right\}\]
\be+\lambda_b |y|+ O\left(
{\kappa ^2, \kappa\lambda_b/M, \lambda_b^2/M^2 } \right)  , \label{rho1}\ee
which is as well symmetric against $y \rightarrow - y$.
One can see that the asymptotic AdS curvature $k$ (defined in the limit $~{y \gg 1/M}$ when $\rho (y) \sim k y$) is somewhat different in the $\tau$ symmetry unbroken and broken phases,
\ba &&k_{unbroken} = \frac23 \kappa M + \lambda_b\quad\mbox{vs.}\no &&k_{broken} = \frac23 \kappa M \Big(1 + \frac{\mu^2}{2M^2}\Big)\sqrt{1 - \frac{\mu^2}{M^2}}+ \lambda_b,\no&&k_{broken}< k_{unbroken}. \label{asymp}
\ea
As the scalar potential is invariant under reflections $ \Phi(y) \longrightarrow - \Phi(y)$ and $  H (y)\longrightarrow -   H(y)$ one finds replicas of the kink-type solutions with different signs of asymptotics at large $y$.
Let us choose further on the positive signs of $ \Phi(y), H(y)$ at $y\to +\infty$ .

 As it was shown in \cite{aags1,partnuc} the second solution with $H(y) \not= 0$ generates  fermion masses whereas the first type of kink with $H(y)= 0$ leaves fermions massless. This is why the second solution is of main interest for our model building. In the fermion lagrangian \cite{aags1,partnuc} the second solution  breaks the $\tau$ symmetry.  Thus there are two phases with different scalar field backgrounds separated by a critical point.  It can be shown (in the zero gravity and thin-brane tension limit) that if  $\Delta_H < M^2/2$ the first kink  provides a local minimum but for some $M^2/2 < \Delta_H < M^2$ it gives a saddle point whereas the second kink with $  H\not= 0$ guarantees a local stability. When gravity and a thin-brane tension is taken into account the position of phase transition point deviates from $M^2/2 = \Delta_H$ (see below).

\subsection{\label{sec:model_nextapprox}Next approximation in $\kappa$:  phase with $\langle H\rangle \neq 0$}

To reduce the complexity of the analytic calculations it is useful to introduce new dimensionless coordinate for the extra dimension,
\be
\tau=M\beta y.
\ee
We are interested in the phase with nonzero $\langle H\rangle$ as it allows to provide the fermions with mass. It can be studied with the help of the perturbation theory in the parameters $\kappa$ expressing the strength of gravity, $\mu/M$ parameterizing the deviation from the critical point and  $\lambda_b/M$ controlling the brane tension.
\ba
&& \Phi(\tau) = M\sum^\infty_{l,m,n=0}\kappa^l \Bigl(\frac{\lambda_b}{M}\Bigr)^m \Bigl(\frac{\mu}{M}\Bigr)^{2n}\Phi_{l,m,n}(\tau),\\
&&H(\tau) = M\sum^\infty_{l,m,n=0}\kappa^l \Bigl(\frac{\lambda_b}{M}\Bigr)^m \Bigl(\frac{\mu}{M}\Bigr)^{2n+1}H_{l,m,n}(\tau),\\
&& \rho(\tau) = \kappa \sum^\infty_{n,m=0}\kappa^n \Bigl(\frac{\mu}{M}\Bigr)^{2m}\rho_{n+1,m}(\tau),\ea \ba
&&\Delta_H
  = \Delta_{H,c}(\kappa) + \frac12 \mu^2,\\
   &&\Delta_{H,c}(\kappa) =\frac12 M^2 \sum^\infty_{m,n=0}\kappa^m\Bigl(\frac{\lambda_b}{M}\Bigr)^n  \Delta_H^{m,n},\\
  &&\Phi_{n,0,0}\equiv\Phi_{n},\quad H_{n,0,0}\equiv  H_{n},\quad\rho_{n,0}\equiv  \rho_{n},\\
&&\frac{1}{\beta^2} = \sum^\infty_{l,m,n=0}\kappa^l \Bigl(\frac{\lambda_b}{M}\Bigr)^m \Bigl(\frac{\mu}{M}\Bigr)^{2n}
\Bigl(\frac{1}{\beta^2}\Bigr)_{l,m,n};
\ea
In \ci{partnuc} we obtained the corrections for background solutions of next order in  $\kappa$ without thin brane,
\be
\Phi_{1,0,0}=-\frac{2}{9}\frac{\sinh{\tau}}{\cosh^3{\tau}},\ee
\ba
H_{1,0,0}&=&\frac{2}{27 \cosh\tau}\no
&&\cdot\left(C^H_{1,0,0} - 2\log\cosh\tau + 3\tanh^2\tau\right),\label{SolutCorr1}
\ea
\be
\Delta_H^{1,0}=-\frac{44}{27},\quad \Bigl(\frac{1}{\beta^2}\Bigr)_{1,0,0} =\frac43,\label{SolutCorr2}
\ee
In the presence of the brane with nonzero tension the background fields get supplementary corrections,
\be
\Phi_{0,1,0}=\epsilon(\tau)\tilde{\Phi}_{0,1,0}(|\tau|),\quad H_{0,1,0}=\tilde{H}_{0,1,0}(|\tau|)
\ee
\be
\Delta_H^{0,1}=-\frac{4}{3}(1+2\ln{2}),\quad \Bigl(\frac{1}{\beta^2}\Bigr)_{0,1,0}=2,
\ee
where $\epsilon(\tau)$ is the sign function,
\ba
\tilde{\Phi}_{0,1,0}&=&-\frac{1}{9}\frac{\tanh\tau}{\cosh^2\tau}\Bigl(9\cosh^2\tau+6\cosh^4\tau\no
&&-6\sinh\tau\cosh\tau^3-12\cosh\tau\sinh\tau\Bigr),
\ea
\ba
\tilde{H}_{0,1,0}&=\frac{1}{3\cosh\tau}\Bigl[&5(\cosh^2\tau-\cosh\tau\sinh\tau)\no
&&+4(\ln\cosh\tau-\tau)\cosh\tau\sinh\tau\no
&&+4(\cosh^2\tau-\tau)\ln{2}+4\tanh\tau\no
&&+\tau+2{\rm Li}_2(-e^{-2\tau})+C^H_{0,1,0}\Bigr]
\ea
and ${\rm Li_2}$ is the dilogarithm function \ci{Lewin},
\be
{\rm Li_2}(z)=-\int_0^z\frac{\ln(1-\zeta)}{\zeta}d\zeta.
\ee
We stress that with these corrections the first derivatives of background fields $\Phi_{1,0}'$ and $H_{1,0}'$ remain continuous.

Mixed orders are practically irrelevant as in realistic models $\kappa \sim 10^{-15}$ and $\mu^2/ M^2 \sim  10^{-3}$ (see \cite{aags2,partnuc}). Correspondingly, $\kappa\mu^2/ M^2 \ll \kappa \ll \mu^2/ M^2$. Therefore the combination of classical solutions \eqref{zeroap}, \eqref{rho1} with the corrections \eqref{SolutCorr1},\eqref{SolutCorr2} provides our calculations with required precision in the case when the perturbation expansion works well. The latter seems to be flawless for classical EoM.

The only problem is that the integration constants $C^H_{1,0,0}$ and $C^H_{0,1,0}$ are not fixed at the first orders in $\kappa, \mu$. To obtain them one
have to study the mixed order approximation of $\Phi$ and $H$ background solutions. It is possible to do analytically in the case without a thin brane. The equation on $\Phi_{1,0,1}$ takes the form,
\ba
(\partial_\tau^2+2-6\Phi_{0}^2)\Phi_{1,0,1}&=&-2\Phi_{1,0,0}+4\Phi_{0}H_{0}H_{1,0,0}\no
&&+6\Phi_{0}^2\Phi_{1,0,0}+2H_{0}^2\Phi_{1,0,0}\no
&&+\frac83 H_{0}^2\Phi_{0}+4\rho_{1,0,1}'\Phi_{0}'\no
&&-2\Bigl(\frac{1}{\beta^2}\Bigr)_{1,0,1}\Phi_{0}(1-\Phi_{0}^2)
\ea
and has a solution,
\ba
&&\Phi_{1,0,1}=\frac{1}{27}\frac{{\rm Li}_2(-e^{-2\tau})-{\rm Li}_2(-e^{2\tau})}{\cosh^2\tau}+\frac{2}{9}\frac{\tanh{\tau}}{\cosh^2{\tau}}\no
&&-\frac{\tau}{\cosh^2{\tau}}\left[\frac{26}{27}+\frac{2}{27}C^H_{1,0,0}+\frac{4}{27}\ln{2}-\frac{1}{2}\Bigl(\frac{1}{\beta^2}\Bigr)_{1,1}\right]
\ea

The equation on $H_{1,0,1}$ looks as follows,
\ba
&&(\partial_\tau^2+1-2\Phi_{0}^2)H_{1,0,1}=-2H_{1,0,0}(1-\Phi_{0}^2)+\frac{8}{27}H_{0}\no
&&-\Bigl(\frac{1}{\beta^2}\Bigr)_{1,0,1}H_{0}(1-2\Phi_{0}^2)
+4H_{0}\Phi_{0}(\Phi_{1,0,1}+\Phi_{1,0,0})\no
&&+6H_{0}^2H_{1,0,0}+\frac83 H_{0}^3+4\rho_{1,0,1}'H_{0}'
\ea

Assuming that $H$ is normalizable, multiplying  this equation by $H_{0}$ and integrating it over $\tau$ from $-\infty$ to $+\infty$ we finally obtain \be C^H_{1,0,0}=-7-2\ln{2}.\ee

Numerical computation gives,
\be C^H_{0,1,0}=+1.3081.\ee


\section{\label{sec:fluct}Field fluctuations around the classical solutions}
Let us consider small localized deviations of the fields
from the  background values and find the action squared in these fluctuations.
In this paper we only outline the derivation of the equations for scalar fluctuations.
The detailed computation can be found in \ci{partnuc}.

The fluctuations of the metric
$ h_ {AB} \left (X \right) $ and of the scalar fields
$ \phi \left (X \right) $ and $\chi \left (X \right) $
 against the background solutions of EoM are introduced in the following way,
    \ba
g_{AB} \left( X \right) dx^A dx^B&=&e^{-2\rho(y)}
\eta _{\mu\nu}dx^\mu dx^\nu-dy^2\no
&&+e^{-2\rho(y)}h_{AB}\left( X \right)dx^A dx^B;\ea
\ba\Phi \left( X \right) =
\Phi \left( y \right) + \phi \left( X \right);\,
H \left( X \right) = H \left( y \right) +
\chi \left( X \right) \label{conf}. \ea

Since 4dim Poincare symmetry is not broken, we select the corresponding
4dim part of the metric $ h_ {\mu \nu} $ and employ the notation for
gravi-vectors $ h_ {5 \mu} \equiv v_ \mu $ and gravi-scalars
$ e^{-2\rho} h_ {55} \equiv S $. The major simplification can be achieved by  separation of
different spin components of the field $ h_ {\mu \nu} $ and $v_\mu$. It  can be
accomplished by description of ten components of
4-dim metric in terms of the traceless-transverse tensor, vector and
scalar components \cite{rev15,bar},
\ba&& h_ {\mu \nu} = b_ {\mu \nu}
+ F_ {\mu, \nu} + F_ {\nu, \mu} + E_ {, \mu \nu} + \eta_ {\mu \nu}
\psi,\no
&&v_\mu=v_\mu^\perp+\partial_\mu\eta, \label {deco} \ea where $ b_ {\mu \nu } $ and $ F_ \mu $ obey
the relation
\be
b_ {\mu \nu} ^ {, \mu} = b = 0 = F_ {\mu} ^ {, \mu}=v_{\mu}^{\perp,\mu}.
\ee After this separation in the action is performed  the
scalar sector that we are interested in decouples from the fields with higher spins up to quadratic orders in fluctuations.

Still there are redundant degrees of freedom because the action \eqref{1} is invariant under diffeomorphisms.
Infinitesimal diffeomorphisms correspond to the Lie derivative
along an arbitrary vector field $\tilde{\zeta} ^A (X) $, defining the
coordinate transformation $ X \to   X = X +  \tilde{\zeta} \left (X \right)$.

If a thin brane defect is switched on in our model we treat it as  partially breaking the gauge (infinitesimal diffeomorphism) symmetry. Particularly to preserve the position of the brane at $y=0$ one has to restrict the allowed diffeomorphism to $\tilde{\zeta}_5|_{y=0}=0$.

Let us rescale the vector fluctuations
$   \tilde\zeta _ \mu = e^{-2\rho} \zeta _ \mu $ and the scalar ones
$   \tilde\zeta _5 = \zeta _5 $ and separate  longitudinal and transverse components in the former,
\be \zeta_ \mu
= \zeta_ \mu ^ \perp + \partial_ \mu C, \qquad \partial ^ \mu \zeta_
\mu ^ \perp = 0.\ee
Then the diffeomorphisms correspond to the following infinitesimal gauge transformation
of the action quadratic in scalar fluctuation
\ba \eta \rightarrow \eta - e^{2\rho}\zeta_5
- C ' , &\quad& E \rightarrow E - 2C, \no \psi \rightarrow \psi +
2\rho' \zeta_5, &\quad& S \rightarrow S - 2 \zeta'_5,\no
\phi \rightarrow \phi + \Phi ' \zeta_5
, &\quad& \chi \rightarrow \chi + H ' \zeta_5.\label{10}
\ea

The further analysis of the scalar spectrum is conveniently performed in the following gauge invariant variables:
\ba \check{\eta}=E' - 2\eta + \frac{e^{2\rho}}{\rho'} \psi; &\qquad& \check{\phi}=\phi+\frac{\Phi'}{2\rho'}\psi;\no \check{\chi}=\chi+\frac{H'}{2\rho'}\psi;&\qquad& \check{S}=S-\frac{1}{\rho'}\psi'+\frac{\rho''}{(\rho')^2}\psi. \label{invvars} \ea
It is important to stress that these variables can have discontinuities as against to initial
non-invariant variables. However because they come from nonzero values of $\psi$ and $\psi'$ allowed
discontinuities in the different variables depend on each other and thus can be restricted.

On the one hand in the case $\lambda_b=0$ the vanishing $\rho'$ means that there can be singularities in $\check{\eta},\check{\phi},\check{\chi}\sim\frac{1}{y}$ and $\check{S}\sim\frac{1}{y^2}$. However such
singularities happen to be forbidden by the integrability of the action.

On the other hand in the case of the nonzero $\lambda_b$ the possible discontinuities in other variables
introduce a $\delta$-singularity into $\check{S}$ again destroying the integrability. Thus the only discontinuities
in the $\check{S}$ are allowed.

The scalar part of the lagrangian quadratic in fluctuations takes the form:
\ba &&\sqrt{\left| g \right|}{\cal L}_{(2),scal}=
\frac12 Ze^{-2\rho}\ \Biggl\lbrace \check{\phi}_{,\mu} \check{\phi}^{,\mu} +
\check{\chi}_{,\mu} \check{\chi}^{,\mu} -\no
&&- e^{-2\rho}\Bigg[ (\check{\phi}')^2 + (\check{\chi}')^2+
\begin{pmatrix}\check{\phi}\\\check{\chi}\end{pmatrix}^T\partial^2 V\begin{pmatrix}\check{\phi}\\\check{\chi}\end{pmatrix}\no
 &&+\frac{1}{2}V(\Phi, H) \check{S}^2 + \check{S}\Bigl(\Phi' \check{\phi}' +
 H' \check{\chi}' -\frac{\partial V}{\partial\Phi} \check{\phi} -
 \frac{\partial V}{\partial H} \check{\chi} \Bigl)\Bigg]\Biggr\rbrace\no
 && + \frac{3}{4} M^3_\ast e^{-4\rho} \square \check{\eta}
  \Bigl(-\rho' \check{S}+\frac{2Z}{3M^3_\ast} (\Phi' \check{\phi} +
 H' \check{\chi} )\Bigr), \label{decoup1} \ea
where $\partial^2V$ is a matrix of second derivatives of the background solutions.

From the last line it follows that the
scalar field $\check{\eta}$ is a gauge invariant Lagrange multiplier and therefore it
generates the gauge invariant
constraint,
\ba \label{cond1} \rho'\check{S} = \frac{2Z}{3M^3_\ast}(\Phi'\check{\phi}+H'\check{\chi}). \ea
Thus after resolving this constraint only two independent
scalar fields remain. To normalize kinetic terms the fields should be
redefined $\check{\psi}=\Omega\hat{\psi}$, $\check{\chi}=\Omega\hat{\chi}$,
where $\Omega=Z^{-1/2}e^{\rho}$. Then the scalar action is reduced to the following form,
\ba
&&\sqrt{\left| g\right|}{\cal L}_{(2),scal}=\frac{1}{2}\Bigl(\partial_{\mu}\hat{\phi}\partial^{\mu}\hat{\phi}+\partial_{\mu}\hat{\chi}\partial^{\mu}\hat{\chi}-\no
&&-e^{-2\rho}\begin{pmatrix}\hat{\phi}\\\hat{\chi}\end{pmatrix}^{T}\Big(-\partial_y^2+2\rho'\partial_y+\hat{{\cal M}}\Big)\begin{pmatrix}\hat{\phi}\\\hat{\chi}\end{pmatrix}\Bigr),\label{flucact}
\ea
 where
 \be
 \hat{{\cal M}}=\partial^2V+\hat{{\cal M}}_{NP}-\rho''+3(\rho')^2
 \ee
 \be
 \hat{{\cal M}}_{NP}=\frac{2Z}{3M_\ast^3}(-\partial_y+4\rho')\left[\frac{1}{\rho'}
 \begin{pmatrix}(\Phi')^2&\Phi'H'\\\Phi'H'&(H')^2\end{pmatrix}\right],\label{massNP}
 \ee
 is a correction to the mass operator that generally speaking can change the spectrum of the scalar fluctuations non-perturbatively.

Let us perform the mass spectrum expansion,
\ba
&&\begin{pmatrix}\hat{\phi}(X)\\\hat{\chi}(X)\end{pmatrix}=
e^\rho\sum_m\Psi^{(m)}(x)\begin{pmatrix}\phi^{(m)}(y)\\\chi^{(m)}(y)\end{pmatrix},\no
&&\partial_{\mu}\partial^{\mu}\Psi^{(m)}=-m^2\Psi^{(m)},
\ea
where the factor $\exp(\rho)$ is introduced to  eliminate first derivatives
in the equations. We obtain the following equations,
\be
\Big(-\partial_y^2+\hat{{\cal M}}-\rho''+(\rho')^2\Big)\begin{pmatrix}\phi^{(m)}\\\chi^{(m)}\end{pmatrix}=
e^{2\rho}m^2\begin{pmatrix}\phi^{(m)}\\\chi^{(m)}\end{pmatrix},\label{flucspec}
\ee
These coupled channel equations of second order in
derivative contain the spectral parameter $m^2$ as 
a coupling constant of a part of potential (a non-derivative piece).
The latter part is essentially negative for all $m^2 >0$. Then as
the exponent $\rho(y)$ is positive and growing at very large $y$ it
becomes evident that the mass term in the potential makes it unbounded
below. Thus any eigenfunction of the spectral problem \eqref{flucspec}
is at best a resonance state though it could be quasi-localized in a
finite volume around a local minimum of the potential. In \cite{aags2}
the probability for quantum tunneling of quasi-localized light resonances
with masses $m\ll M$ was estimated as $\sim \exp\{-\frac{3}{\kappa}\ln\frac{2M}{m}\}$
which for phenomenologically acceptable values of $\kappa \sim 10^{-15}$ and
$M/m \gtrsim 30$ means an enormous suppression.
Moreover in the perturbation theory the decay does not occur as the turning
point to an unbounded potential energy is situated at $y\sim 1/\kappa$.
Therefore one can calculate the localization of resonances  following the perturbation schemes.

To obtain the spectrum we also have to consider the boundary terms taking into
account Gibbons-Hawking supplement,
\be
S_{(2),scal}^{(bound)}=-3M_\ast^3\int\limits_{bound}d^4x\sqrt{g^{(4)}}\left[\rho'\check{S}^2\right]_\pm
\ee
Combining boundary terms arising in variations of $\check\phi, \check\chi$ we obtain the
following matching conditions,
\ba
&&\left[\rho'\check{S}\right]_{\pm}=0,\no
&&\left[2\partial_y\check{\phi}+\Phi'\check{S}\right]_{\pm}=0,\quad
\left[2\partial_y\check{\chi}+H'\check{S}\right]_{\pm}=0.
\ea
The first condition ensures that the constraint \eqref{cond1} can be applied on both
sides of the brane without discontinuities of $\check{\phi}$ and $\check{\chi}$.
After solving the constraint the matching conditions for profile functions look as follows,
\ba
\left[\partial_y\phi^{(m)}\right]_{\pm}&=&-\frac{2Z}{3M^3_\ast}\frac{\Phi'|_{y=0}}{\rho'|_{0+}}\left(\Phi'\phi^{(m)}+H'\chi^{(m)}\right)\Big|_{y=0}\no
&&-4\rho'|_{0+}\phi^{(m)},\label{matchingphi}\\
\left[\partial_y\chi^{(m)}\right]_{\pm}&=&-\frac{2Z}{3M^3_\ast}\frac{H'|_{y=0}}{\rho'|_{0+}}\left(\Phi'\phi^{(m)}+H'\chi^{(m)}\right)\Big|_{y=0}\no
&&-4\rho'|_{0+}\chi^{(m)}\label{matchingchi}
\ea

To define the limit of switched-off gravity it is useful to parameterize the interaction
with the parameter $\kappa$ in a way similar to the minimal model $Z=\frac{3\kappa M_\ast^3}{M^2}$
where $M$ is a characteristic scale of the scalar field interaction.  If there is no thin
brane $\lambda_b=0$ the spectral equation \eqref{flucspec} gets nontrivial corrections in the
limit $\kappa\longrightarrow 0$ because
\be
 \hat{{\cal M}}_{NP}\longrightarrow-\frac{2}{M^2}\partial_y\left[\frac{\kappa}{\rho'}
 \begin{pmatrix}(\Phi')^2&\Phi'H'\\\Phi'H'&(H')^2\end{pmatrix}\right]\neq 0,
\ee
Similarly eqs. \eqref{matchingphi},\eqref{matchingchi} remain nontrivial because of the
factor $\kappa/\rho'$. As we will see in the next section the latter eqs.  drastically change the
spectrum of the scalar fluctuations.

However if one introduces a sufficiently large positive brane tension $\lambda_b\gg\kappa M$
these singular effects wash out completely both in the spectral equation outside the brane and in the
matching conditions. Thus one can organize a smooth limit to the model without gravity keeping the brane tension large in comparison with $\kappa M$. It is interesting to compare this result with the limit of $\kappa\rightarrow 0$ when the brane tension is coherently $\lambda_b\simeq\kappa M$ and the singular effects remain nontrivial.

\section{\label{sec:phichannel}Fluctuations in the $\phi$-channel around a background with $\langle H\rangle=0$}
\subsection{\label{sec:phichannel_eq}The equation and the matching condition}

When $\langle H\rangle=0$ eqs. of motion \eqref{eoms} entail $\partial^2V/\partial\Phi\partial H=0$
 and the two scalar sectors decouple. In this case
  the equation on $\phi^{(m)}$ \eqref{flucspec} can be written using eqs. \eqref{eoms0}-\eqref{eoms} in the following factorized form,
\ba
\left(-\partial_y+{\cal P}\right)
\left(\partial_y+{\cal P}\right)\phi^{(m)}=e^{2\rho}m^2\phi^{(m)},\label{branoneq}
\ea
where ${\cal P}=\frac{\rho''}{\rho'}-\frac{\Phi''}{\Phi'}+2\rho'$.

This equation aside the brane should be supplemented with the matching condition,
\ba
\left[\partial_y\phi^{(m)}\right]_{\pm}=
-\frac{2Z}{3M^3_\ast}\frac{(\Phi'|_{y=0})^2}{\rho'|_{0+}}\phi^{(m)}|_{y=0}\no
-4\rho'|_{0+}\phi^{(m)}.\label{matching}
\ea
One may interpret this matching condition as the contribution of a $\delta$-potential,
\be
-\left(\frac{Z}{3M^3_\ast}\frac{(\Phi'|_{y=0})^2}{\rho'|_{0+}}
+2\rho'|_{0+}\right)\delta(y).
\ee

If assuming that $\lambda_b=\kappa M b$ in the zero
gravity limit $\kappa\rightarrow 0$ the solutions in the $\phi$ channel  can be obtained exactly. Eq. \eqref{branoneq}
in this limit can be written as,
\ba
&Q_bQ_b^\dagger\phi^{(m)}=\frac{m^2}{M^2}\phi^{(m)},\\
&Q_b=-\partial_\tau+\frac{\tilde{\rho}_1''}{\tilde{\rho}_1'+b}-\frac{\Phi''}{\Phi'},\\ &Q_b^\dagger=\partial_\tau+\frac{\tilde{\rho}_1''}{\tilde{\rho}_1'+b}-\frac{\Phi''}{\Phi'},
\ea
where $\tau=M\beta y$ and $\rho=\kappa\tilde{\rho}_1+\kappa b \tau+O(\kappa^2)$.

Generally speaking the potential may have singularities where $\tilde{\rho_1'}+b$ vanishes.
One can solve this equation in the regions where the potential is regular and match the solution
appropriately at the singular points and on the brane. We can use the factorization of the
potential to construct the super-partner potential \cite{susy1,susy2} that happens  not to depend on $b$,
\ba
&&Q_b^\dagger Q_b=-\partial_\tau^2+4-\frac{2}{\cosh^2\tau}.
\ea

The solutions with $m^2\neq 0$ can be constructed from the solutions for the
super-partner potential,
\be
\phi^{(m)}=Q_b\check{\phi}^{(m)},\quad Q_b^\dagger Q_b\check{\phi}^{(m)}=
\frac{m^2}{M^2}\check{\phi}^{(m)}
\ee

The super-potential can be factorized in an adjacent way that connects its
solution with the solutions of the constant potential,
\be
\tilde{Q}=-\partial_\tau+\tanh\tau,\quad \tilde{Q}=\partial_\tau+\tanh\tau
\ee
\be
Q_b^\dagger Q_b = \tilde{Q}\tilde{Q}^\dagger + 3,\quad \tilde{Q}^\dagger\tilde{Q}=1
\ee

This allows us to obtain the solutions of the continuous spectrum,
\ba
&&f^{(m)}_b=Q_b\tilde{Q}\sin{k\tau},\quad g^{(m)}_b=Q_b\tilde{Q}\sin{k\tau}\cos{k\tau},\no
&& m^2=M^2(4+k^2)
\ea
Their linear combination vanishing at $y=0$ reads,
\ba
&& f^{(m),0}_b=b(1+k^2)f^{(m)}-kg^{(m)},\no
&& (f^{(m),0}_b)'|_{y=0}=-k(k^2+1)(k^2+4)b.\label{PhiSolWith0}
\ea

Both solutions can also be considered for imaginary $k$ corresponding to the mass lower
than $2M$. Without matching it is possible to construct the solution decreasing at infinity
 on  one side of the brane however it  exponentially increases on another side.
 With the help of  matching it may be possible to construct a normalizable solution with
 the matching conditions restricting the values of $k$.

The spectrum of the potential $Q_b^\dagger Q_b$ also contains the states with $m^2=3M^2$
that cannot be retrieved from the constant potential. For the branon potential it gives two
possible solutions,
\ba
&&\phi^{(\sqrt{3}M)}_b=Q_b\frac{1}{\cosh\tau},\\
&&\tilde{\phi}^{(\sqrt{3}M)}_b=Q_b\Bigl(\sinh\tau+\frac{x}{\cosh\tau}\Bigr).
\ea
The first one is decreasing at infinity while the second one is increasing exponentially.

Besides the solutions constructed from the super-potential states  can also happen to
be zero-modes. They can be written in the model-independent way,
\be
\omega_b=\frac{\Phi'}{\rho'},\quad
\tilde{\omega}_b=\omega_b\int^\tau d\tau'\frac{1}{\omega_b^2(\tau')}
\ee
Notice that $\omega_b$ may have singularities where $\rho'$ vanishes but decreases at
infinity and vice versa $\tilde{\omega}_b$ vanishes simultaneously with $\rho'$ but
asymptotically goes to infinity. With appropriate boundaries they still can contribute
to the fluctuations.

\subsection{\label{sec:phichannel_nobrane}The spectrum in the absence of a thin brane}

In this case the potential has a singular barrier $\sim\frac{2}{\tau^2}$ preventing
the appearance of localized solutions \cite{rev19}. There exist two sets of solutions
of the continuous spectrum corresponding to the particles living on different sides of
the brane.
\be
\phi^{(m)}_{>}=
\begin{cases}
    g^{(m)}_0,& \tau> 0\\
    0,              & \tau<0
\end{cases},
\quad
\phi^{(m)}_{<}=
\begin{cases}
    0,& \tau> 0\\
    g^{(m)}_0,              & \tau<0
\end{cases},
\ee
where
\ba
g^{(m)}_0&=&\frac{3k^2-(4+k^2)\tanh^2\tau}{3-\tanh^2\tau}\cos{k\tau}\no
&&+\frac{3k(1+\tanh^2\tau)}{\tanh\tau(3-\tanh^2\tau)}\sin{k\tau},
\ea
with masses $m^2=M^2(4+k^2)$. Both $g^{(m)}_0$ and its first derivative vanish at
$\tau=0$ . Therefore $\phi^{(m)}_{>}$ and $\phi^{(m)}_{<}$ satisfy the matching
conditions.

Concerning the localized solutions both $\phi^{(\sqrt{3}M)}_0$ and $\omega_0$ happen
to be singular and $\tilde{\omega}_0$ is not normalizable. Also there is no state
constructed from the solutions with imaginary $k$. Thus there is no localized
solution in this case including a (normalizable) Goldstone zero-mode related to
spontaneous breaking of translational
symmetry. The reason is evident: the corresponding brane fluctuation represents, in
fact, a gauge transformation \eqref{10} and does not appear in the invariant part
of the spectrum. One could say that in the presence of gravity induced by a brane
the latter becomes  more rigid as only massive fluctuations are possible around it.
Of course, the very gauge transformation \eqref{10} leaves invariant only the
quadratic action and  thereby a track of Goldstone mode may have influence
on higher order vertices of interaction between gravity and scalar fields.
This option is beyond the scope of the present investigation.

\subsection{\label{sec:phichannel_posbrane}The spectrum in the case of the positive brane tension}

In this case the singular potential is shifted to the region which is cut
off after matching.

Again there exist two sets of solutions of the continuous spectrum. The first set
is composed of the solutions $g^{(m)}_b$ with different signs of $b$ on different
sides of the brane
\ba
&&\phi^{(m)}=
\begin{cases}
    g^{(m),0}_b,& \tau> 0\\
    g^{(m),0}_{-b},              & \tau<0
\end{cases},\quad m^2=M^2(4+k^2), \no
&&\phi^{(m)}|_{y=0}=-(k^2+1),\no
&&(\phi^{(m)})'|_{y=0+}=(\phi^{(m)})'|_{y=0-}=\frac{k^2+1}{b},
\ea

The second set of solutions from continuous spectrum is constructed from matching
the solutions \eqref{PhiSolWith0} vanishing at $\tau=0$ which have different signs of $a$ on
different sides of the brane,
\ba
&&\tilde{\phi}^{(m)}=
\begin{cases}
    f^{(m),0}_b,& \tau> 0\\
    f^{(m),0}_{-b},              & \tau<0
\end{cases},\quad m^2=M^2(4+k^2), \no
&&\tilde{\phi}^{(m)}|_{y=0}=0,\no
&&(\tilde{\phi}^{(m)})'|_{y=0+}=
(\tilde{\phi}^{(m)})'|_{y=0-}
\ea

The zero-mode takes the form,
\ba
&&\phi^{(0)}=
\begin{cases}
    \omega_b,& \tau> 0\\
    -\omega_{-b},& \tau<0
\end{cases},\no
&&\phi^{(0)}|_{y=0}=\frac{\Phi'(0)}{b},\no
&&(\phi^{(0)})'|_{y=0+}=
-(\phi^{(0)})'|_{y=0-}=-\frac{1}{3b^2},,
\ea
 and satisfies the matching conditions \eqref{matching}.

The possible heavy localized state $\phi^{(\sqrt{3}M)}_b$ as a symmetric solution
does not satisfy the matching conditions \eqref{matching} and as an antisymmetric
solution has a discontinuity $[\phi^{(\sqrt{3}M)}_b\Bigl]_{\pm}=2/b$. Thus there
is no a localized state with this mass. Yet it is possible to construct the massive
localized state from the continuous spectrum solutions with imaginary $k$. Such a state is given by,
\ba
&&\phi^{(m_h)}=
\begin{cases}
    f^{(m_h),0}_b,\\
    -f^{(m_h),0}_{-b}
\end{cases},
\quad k_h=i\tilde{k}_h=i\frac{-1+\sqrt{1+4a^2}}{2a},\no
&&\quad m_h^2=M^2\frac{-1+\sqrt{1+4a^2}+6a^2}{2a^2}
\ea
\ba
f^{(m_h),0}_b&=&\frac{e^{-\tilde{k}_h\tau}}{\cosh^2\tau}
\frac{1}{3\tanh\tau-\tanh^3\tau+3b}\no
&&\cdot\Big[\tilde{k}_h^2(1+2\cosh^2\tau)+\sinh^2\tau\no
&&\quad\qquad\qquad+3\tilde{k}_h\sinh\tau\cosh\tau\Big]
\ea

In the limit of very large $b$ the potential become the potential of the model without gravity,
\be
Q_bQ_b^\dagger\rightarrow-\partial_\tau^2+4-\frac{6}{\cosh^2\tau}
\ee
and the matching conditions become trivial. The
zero-mode and a massive localized state in this limit coincide with the corresponding states
in the model without gravity,
\ba
&&b\phi^{(0)}=\rightarrow \frac{1}{\cosh^2\tau}=\Phi',
\quad b\phi^{(m_h)}\rightarrow \frac{\sinh\tau}{\cosh^2\tau},\no
&& m_h^2\rightarrow 3M
\ea

On the other hand in the limit $b\rightarrow 0$ the zero-mode becomes singular and the
corresponding action becomes non-integrable. The massive localized state in this limit coincides
with the continuous spectrum threshold state $g^{(0)}_0$ with mass $m_h^2\rightarrow 4M^2$.

Let us now consider the limit of vanishing gravity $\kappa\rightarrow 0$ with the brane tension
$\lambda_b$ remaining large compared to $\kappa M$. Then the equation \eqref{branoneq} aside the brane combined with matching condition \eqref{matching} can be written as,
\ba
&\Big(&-\partial_\tau^2+\frac{1}{M^2\beta^2}\frac{\partial^2V}{\partial\Phi^2}\no
&&+\frac{4\lambda_b^2}{M^2\beta^2}-\frac{4\lambda_b}{M\beta}\delta(\tau)\Big)\phi^{(m)}=
m^2e^{2\lambda_b|\tau|/M\beta}\phi^{(m)}
\ea

For $\lambda_b\rightarrow 0$ both the equations aside the brane take the form similar to the model without gravity,
\be
\left(-\partial_\tau^2+\frac{1}{M^2\beta^2}\partial^2V\right)\phi^{(m)}=m^2\phi^{(m)},
\ee
and the matching conditions become trivial ($\delta$-function in the potential vanishes.)

\subsection{\label{sec:phichannel_negbrane}The spectrum in the case of the negative brane tension}

We can also consider the case of a small negative brane tension. Notice that
$b>-\frac{2}{3}$ is chosen in order to keep the Anti-de-Sitter geometry at large $\tau$. In this case we obtain two
singular barriers at $\pm \tau_b$, $\tanh^3\tau_b-3\tanh\tau_b-3b=0$
that divide the bulk space-time into three parts essentially separated from each other. The spectrum
outside the well made by these barriers is continuous,
\ba
&&\phi^{(m)}_{>}=
\begin{cases}
    A_af^{(m)}_b+B_bg^{(m)}_b,& \tau> \tau_b\\
    0,              & \tau<\tau_b
\end{cases},
\\
&&\phi^{(m)}_{<}=
\begin{cases}
    0,& \tau> -\tau_b\\
    A_{-b}f^{(m)}_{-b}+B_{-b}g^{(m)}_{-b},              & \tau<-\tau_b
\end{cases},
\ea
where
\ba
&&A_b=\frac{k\sin{k\tau_b}\cosh{\tau_b}+\sinh{\tau_b}\cos{k\tau_b}}{\cosh^5 \tau_b},\\ &&B_b=\frac{k\cos{k\tau_b}\cosh{\tau_b}-sin{k\tau_b}\sinh{\tau_b}}{\cosh^5\tau_b}
\ea

Inside the well there exists a discrete spectrum of states that vanish at $\tau=\pm\tau_b$ and remain
zero outside the potential well. There exist two sets of these solutions that are constructed by
matching the solutions $f^{(m)}_b$ and $g^{(m)}_b$. The first set is composed of the solutions
$g^{(m)}_b$ with different signs of $b$ on different sides of the brane
\ba
&&\phi^{(m)}=
\begin{cases}
    g^{(m),0}_b,& 0<\tau<\tau_b\\
    g^{(m),0}_{-b},              & -\tau_b<\tau<0\\
    0,                          &|\tau|>\tau_b
\end{cases},\no
&& m^2=M^2(4+k^2), \no
&&\phi^{(m)}|_{y=0}=-(k^2+1),\no
&&(\phi^{(m)})'|_{y=0+}=-(\phi^{(m)})'|_{y=0-}=\frac{k^2+1}{b},
\ea
The condition on $k$ takes the form,
\be
\tan{k\tau_b}=-\tanh\tau_b
\ee
and thus the spectrum is,
\ba
&&k_n=-\frac{\arctan(\tanh\tau_b)}{\tau_b}+\pi n,\quad n\in\mathbb{Z},\no
&&m_n^2=M^2(4+k_n^2)
\ea

The second set of solutions from continuous spectrum is constructed by matching  solutions \eqref{PhiSolWith0} vanishing at
$\tau=0$ and with different signs of $b$ on different sides of the brane,
\ba
&&\tilde{\phi}^{(m)}=
\begin{cases}
    f^{(m),0}_b,& \tau> 0\\
    f^{(m),0}_{-b},              & \tau<0
\end{cases},\quad m^2=M^2(4+k^2), \no
&&\tilde{\phi}^{(m)}|_{y=0}=0,\no
&&(\tilde{\phi}^{(m)})'|_{y=0+}=(\tilde{\phi}^{(m)})'|_{y=0-}.
\ea
The condition on $k$ can be written in the form,
\ba
\tan{k\tau_b}&=&k\frac{(1+k^2)\tanh^3\tau_b-3k^2\tanh\tau_b}{-3k^2-(1+k^2)\big(3\tanh^2\tau_b- \tanh^4\tau_b\big)}\no
&\simeq&\frac{3\tanh\tau_b-\tanh^2\tau_b}{\tanh^4\tau_b-3\tanh^2\tau_b-3}\frac{1}{\tau_b}
\cdot(k\tau_b),
\ea
where $k>0$.

For very small $\tau_b$ it can be written simply as\\ $\tan{\tau_b k}=\tau_b k$ .
As expected for $b\rightarrow 0$, $\tau_b\rightarrow 0$ these states become very heavy and
decouple. This condition on $k$ is satisfied by the two solutions with imaginary $k=i,2i$ that are allowed
because  the function has a compact support by construction. However they happen to be trivial $\phi=0$.

$\omega_b$ is singular at $\tau=\pm\tau_b$. On the other hand both $\tilde{\omega}_b$ and
$\tilde{\omega}_b'$ vanishes there and can be matched with zero outside the well thus constructing
the normalizable zero-mode inside the well. However $\tilde{\omega}$ does not satisfy the matching
conditions \eqref{matching}. The matching condition is not satisfied also by a possible state
constructed from $\phi_b^{(\sqrt{3}M)}$ and $\tilde{\phi}_b^{(\sqrt{3}M)}$. Again one can interpret 
the matching condition as a $\delta$-potential. As opposed to the case of the positive brane tension
now we get not a well but a $\delta$-barrier.

\section{\label{sec:lightscalar}Light scalar mode in different phases and at a critical point}

\subsection{\label{sec:lightscalar_zeroH}Fluctuations around  a background with $\langle H\rangle=0$}

Consider now the fluctuations in the second channel in the phase with $\langle H\rangle=0$,
\be
\left(-\partial_y^2+\frac{\partial^2V}{\partial H^2}+4(\rho')^2-2\rho''\right)\chi_m
=e^{2\rho}m^2\chi^{(m)},\label{higgs}
\ee

These fluctuations of the second, mass generating field $H(y)$ do not develop any singular
barrier and its potential
is regular and for background solutions delivering a minimum this operator
must be positive. For the minimal potential with quartic self-interaction \eqref{36}
one can derive  more quantitative
conclusions. Indeed, for gravity switched off the background $ \Phi(y) = \Phi_0(y)$
is defined by \eqref{kink0} . Accordingly the mass spectrum operator receives the potential
\ba
{\cal V}(y) &=& - 2\Delta_H + 2 \Phi_0^2\no
& =& (M^2 - 2\Delta_H) + M^2\Big(1 - \frac{2}{\cosh^2 My}\Big).
\ea
The only localized state of the mass operator
$\hat m^2_\chi$ is $\hat\chi \rightarrow \chi_0 \simeq 1/\cosh(My)$ with the
corresponding mass $m^2_0 = M^2 - 2\Delta_H$ as expected. Thus in the unbroken
phase with $M^2 > 2\Delta_H$ the lightest scalar fluctuation in $\chi$ channel
possesses a positive mass and the system is stable.
In the critical point,  $M^2 = 2\Delta_H$, a lightest fluctuation is massless and
for $M^2 < 2\Delta_H \leq 2M^2$ the localized state $\chi_0$ represents a tachyon
and brings instability providing a saddle point. Instead the solution \eqref{zeroap}
provides a true minimum (see \cite{aags1}). At critical point this mode is still massless when we
 calculate the next orders in $\kappa$ corrections to the background solutions and to the
parameter $\Delta_{H,c}$ \ci{partnuc} . The condition that the mass of this state is zero happens
to be equivalent to the normalizability of the correction to $\Phi$.  This result is easily 
generalized for the arbitrary value of the thin brane tension $b$.

\subsection{\label{sec:lightscalar_mass}Fluctuations in the phase with $\langle H\rangle\neq0$}

In the broken phase the mixing terms are nonzero and one has to study the scalar spectrum by means of  perturbation
theory near critical point. From now on let us assume  that $\lambda_b=\kappa M b$. The massless state gains  a mass and a nonzero $\phi$ component,
 \ba
&& \chi^{(m)}=\sum_{n,k}^{\infty}\kappa^n\Bigl(\frac{\mu}{M}\Bigr)^k\chi_{n,k},\no
&& \phi^{(m)}=\sum_{n,k}^{\infty}\kappa^n\Bigl(\frac{\mu}{M}\Bigr)^{k+1}\phi_{n,k},\no
&&  m^2=M^2\sum_{n,k}^{\infty}\kappa^n\Bigl(\frac{\mu}{M}\Bigr)^k (m^2)_{n,k},
 \ea

Iteration of the equations \eqref{flucspec} is organised as follows,
 \ba
  &&-Q_bQ_b^\dagger\phi_{0,k}+(m^2)_{k}\phi_{0,0}={\cal F}_{0,k},\label{iterphi}\\
  &&-\tilde{Q}\tilde{Q}^\dagger\chi_{0,k+1}+(m^2)_{k+1}\chi_{0,0}={\cal H}_{0,k}\label{iterchi}
 \ea
 where the definitions of $Q_b$ and $\tilde{Q}$ are the same as in Sect. \ref{sec:phichannel} and
 \ba
 {\cal F}_{0,k}&=&\sum_{l=0}^k\Bigg[4\Phi_{0}H_{0}-
  \left((-1)^{k-l}\frac{2(\rho_{1,1}')^{k-l}}{(\rho_{1,0}')^{k-l+1}}\Phi_{0}'H_{0}'\right)'\Bigg]\chi_l
  \no
  &&+\sum_{l=0}^{k-1}\Bigg[-2+6\Phi_{0}^2+2H_{0}^2\no
  &&\qquad\qquad+\left((-1)^{k-l}\frac{2(\rho_{1,1}')^{k-l+1}}{(\rho_{1,0}')^{k-l+2}}\Phi_{0}^2\right)'\Bigg]\phi_{0,l}\no
  &&-\sum_{l=0}^{k-1}\sum_{r=0}^{k-l}(m^2)_{l}\phi_{0,r},\\
   {\cal H}_{0,k+1}&=&\sum_{l=0}^k\Bigg[4\Phi_{0}H_{0}-
  \left((-1)^{k-l}\frac{2(\rho_{1,1}')^{k-l}}{(\rho_{1,0}')^{k-l+1}}\Phi_{0}'H_{0}'\right)'\Bigg]\phi_l
  \no
  &&+\sum_{l=0}^{k}\Bigg[-2+2\Phi_{0}^2+
  6H_{0}^2\no
  &&\qquad\qquad-\left((-1)^{k-l}\frac{2(\rho_{1,1}')^{k-l+1}}{(\rho_{1,0}')^{k-l+2}}H_{0}^2\right)'\Bigg]\chi_{0,l}\no
  &&-\sum_{l=1}^{k}\sum_{r=0}^{k-l+1}(m^2)_{l}\phi_{0,r}.
 \ea

  The solutions can be obtained by integration of these equations,
\ba
 &\phi_{0,k}(\tau)&=\omega(\tau)\int_0^\tau d\tau'\frac{1}{\omega_b^2(\tau')}
 \int_{-\infty}^{\tau'}d\tau''\no
 &&\quad\omega(\tau'')\left({\cal F}_{0,k}(\tau'')-
 (m^2)_{0,k}\phi_{0,0}(\tau'')\right),\\
&\chi_{0,k}&=\frac{1}{\cosh\tau}\int^\tau d\tau'\cosh^2\tau'
\int_{0}^{\tau'}d\tau''\no
&&\frac{1}{\cosh\tau''}\left({\cal H}_{0,k}(\tau'')
-(m^2)_{0,k}\chi_{0,0}(\tau'')\right),
\ea
where $\omega_b=\frac{\Phi_{0}'}{\rho_{1,0}'}$ is a singular zero-mass
solution in the $\phi$ channel. The mass is obtained by integration of the
equation for the corresponding order of $\chi$ assuming its normalizability,
\be
(m^2)_{0,k}=\frac{\int_{-\infty}^{+\infty}\chi_{0,0}{\cal H}_{0,k}d\tau}{\int_{-\infty}^{+\infty}\chi_{0,0}^2d\tau}.
\ee

Using eqs. \eqref{iterphi},\eqref{iterchi} and the matching conditions \eqref{matchingphi},\eqref{matchingchi} we obtain the next correction to $\phi$ and $\chi$,
\be
\phi_{0,0}(\tau)\Big|_{b\geq 0}=\epsilon(\tau)\tilde{\phi}_{b}(|\tau|),\quad \chi_{0,0}\Big|_{b\geq 0}=\tilde{h}(|\tau|)
\ee
\ba
\tilde{\phi}_b=\frac{\tanh^2\tau-2\ln\cosh\tau-3b\tau}{\cosh^2\tau\cdot\left(3b+3\tanh^2\tau-\tanh^3\tau\right)}
\ea
\ba
\tilde{\chi}_b&=&\frac{C^\chi_{0,1}}{\cosh\tau}+\frac{\tanh\tau}{2\cosh\tau(2\cosh^2\tau+1)+6b\cosh^2\tau}\no
&&\cdot\left[\Bigl(4\cosh^2\tau\ln(2\cosh\tau)+2\ln{2}+3\Bigr)+\right.\no
&&\quad\left.+b\cosh^2\tau\cdot\Bigl(6\tau+(3+6\ln{2})\coth\tau\Bigr)\right]
\ea
where $C^\chi_{0,1}$ depends on the normalization of $\chi^{(m)}$. In order to keep the same normalization  $\chi_{0,1}$ should be
orthogonal to $\chi_{0,0}$. In the case $b=0$ the constant can be found analytically,
\ba
C^\chi_{0,1}\Big|_{b=0}&=&\frac{1}{2\sqrt{3}}\left({\rm Li_2}\Bigl(\frac{2}{1-\sqrt{3}}\Bigr)-
{\rm Li_2}\Bigl(\frac{2}{\sqrt{3}+1}\Bigr)\right)\no
&&-\left(\frac32+\ln{2}\right)\left(1+\frac{1}{\sqrt{3}}\ln\frac{\sqrt(3)-1}{\sqrt(3)+1}\right)\no
&\approx&-1.322
\ea

The leading order of the light scalar mass does not depend on $b\geq 0$ and happens to be the same
as in the model \cite{aags1} without gravity,
\ba
(m^2)_{0,1}\Big|_{b\geq 0}=\frac{\int_{-\infty}^{+\infty}\chi_{0,0}{\cal H}_{0,1}d\tau}{\int_{-\infty}^{+\infty}\chi_{0,0}^2d\tau}=2,
\ea

In the case of negative thin brane tension $b<0$ there are singularities in the potential in the
$\phi$ channel and in the mixing terms at $\tau=\pm\tau_b$. The leading order of $\phi$ is,
\ba
\phi_{0,0}\Big|_{b<0}&=&\epsilon(\tau)\Big(\tilde{\phi}_b(|\tau|)+C^{(\phi)}_{0,0}\omega_b(|\tau|)\Big)\no
&&+\begin{cases}
    \tilde{C}^{(\phi)}_{0,0}\epsilon(\tau)\tilde{\omega}_b(|\tau|)\Big),& |\tau|<\tau_b\\
    0,& |\tau|>\tau_b
\end{cases}
\ea
where
\be
\tilde{\omega}_b=\omega_b\int^\tau d\tau'\frac{1}{\omega_b^2(\tau')}
\ee
and the constants are chosen to subtract singularities at $\tau=\pm\tau_b$ and for $\phi_{0,0}\Big|_{\tau=0}=0$
to satisfy the matching conditions,
\ba
&&C^{(\phi)}_{0,0}=3b\tau_b+2\ln\cosh\tau_b-\tanh^2\tau_b,\\
&&\tilde{C}^{(\phi)}_{0,0}=-C^{(\phi)}_{0,0}\frac{\omega_b(0)}{\tilde{\omega}_b(0)}.
\ea

Notice that while it is possible to choose the integration constants in such a way that there is no
singularity at $\pm\tau_b$ the condition $\phi_{0,0}\Big|_{\pm\tau_b}=0$ cannot be satisfied. At first sight it seems that such a solution is forbidden by the condition of finiteness of the action $\eqref{flucact}$.
However because this is a solution of the spectral equation the singularities of the potential compensate
each other and the integrand happens to be integrable.

The leading order of the mass is given by,
\ba
(m^2&)_{0,1}&\Big|_{b< 0}=\frac{p(\tau_b)}{q(\tau_b)}\\
p(\tau)&=&8\ln c(\tau)\cdot s(2\tau) c^2(\tau)\cdot( s(2\tau)+2\tau( c(2\tau)+2))\no
&&-\frac12 s(2\tau)\cdot(2 c(\tau)- s(\tau))\cdot\Bigl(-3 s(\tau)+3 s(3\tau)\no
&&+8 c(\tau)+4 c(3\tau)\Bigr)+\tau\Bigl(7 s(2\tau)-2s(4\tau)\no
&&-s(6\tau)+9 c(2\tau)+15\Bigr)+4 \tau^2 (3 c(2\tau)+5)-\no
&&-32\left((\ln c(\tau))^2+\tau^2\right)\cdot c^6(\tau)\\
q(\tau)&=&- s(\tau) c(\tau)(5 c^2(\tau)\cdot( c(2\tau)+2)-3)\no
&&+3\tau(3 c^2(\tau)+1)
\ea
where $s(\tau)=\sinh\tau$,$c(\tau)=\cosh\tau$. For $b\longrightarrow 0$ there is a smooth limit $(m^2)_{0,1}\longrightarrow 2$. For $b\longrightarrow-\frac{2}{3}$,
\be
(m^2)_{0,1}\longrightarrow \frac{14}{5}+\frac{16}{5}(\ln{2})^2+\frac{16}{5}\ln{2}\approx 6.5555
\ee
The higher orders can be computed analytically in the case $b=0$. The first order in the expansion of $\phi$ reads,
 \ba
\phi_{0,1}\Big|_{b=0}&=&-(2\ln{2}+1)\epsilon(\tau)\tilde{\omega}_0
-\frac{1}{3s(2\tau)(1+2c^2(\tau))^2}\no
&&\cdot\Bigg[9-8c^6(\tau)+7c^4(\tau)
-4c^2(\tau)-4c^8(\tau)\no
&&\quad+6\left(c^2(\tau)-2c^4(\tau)+1\right)\ln{2}\no
&&\quad+2\ln{c(\tau)}\cdot(6c^2(\tau)\cdot(1+2c^2(\tau))\ln{2}\no
&&\quad+14c^4(\tau)+4c^8(\tau)+6c^6(\tau)+15c^2(\tau))\no
&&\quad+\tau s(2\tau)(-8c^4(\tau)-4c^6(\tau)+2+c^2(\tau))\no
&&\quad+3c^2(\tau)\cdot(1+2c^2(\tau))\tau^2\no
&&\quad+6C^\chi_{0,1}\Bigg(c^2(\tau)-2c^4(\tau)+1\no
&&\quad+2c^2(\tau)
\cdot(1+2c^2(\tau))\ln{c(\tau)}\Bigg)\Bigg],
\ea
where $\epsilon(\tau)$ is a sign function and
\ba
\tilde{\omega}_0&=&\omega_0\int^\tau d\tau'\frac{1}{\omega_0^2(\tau')}\no
&=&\frac{1}{6\sinh\tau(2\cosh^2\tau+1)}\cdot\Big(2\sinh\tau\cosh^4\tau\no
&&\:\:+
3\sinh\tau\cosh^2\tau-3\tau\cosh\tau-2\sinh\tau\Big)
\ea
While $\phi_{0,1}$ is continuously differentiable its second derivative
has discontinuity at $\tau=0$ behaving near that point as
$\sim\pm(\frac23\ln{2}+\frac13)\tau^2$. The discontinuity is compensated
in the equation by terms with $\phi_{0,1}$ multiplied on singularity of
the potential $\sim{\rm const}/\tau^2$.

The next-to-leading order of the mass expansion looks as follows,
\ba
(m^2)_{0,2}\Big|_{b=0}&=&-128\sqrt{3}{\rm arctanh}\frac{\sqrt{3}}{3}+146\no
&&+\frac43\ln{2}\cdot(1+\ln{2})-\frac{\pi^2}{9}\no
&\approx& +0.4817,
\ea
whereas the analogous computation for a similar model without gravity gives,
\be
(m^2)_{0,2}^{NG}=-\frac{130442}{121275}\approx -1.0756 .
\ee

Thus we have revealed the unambiguous puzzle of discontinuity in the mass spectrum of scalar fluctuations
 between a theory defined without gravity and a theory
with minimal gravity interaction in the zero gravity limit.

 \section{\label{sec:conclusions}Summary and outlook}
In this paper we have thoroughly calculated the gravity effects on the formation of a thick brane and on the localization of light scalar states  including the space-time defect in the form of the rigid thin brane with a small tension. This defect allows to understand better the puzzle of non-analytical limit of vanishing gravity. As well in phenomena of spontaneous breaking of space-time symmetries tiny defects point out the true position where domain walls (branes) are generated. The  results obtained in the paper include:
\begin{itemize}
\item The corrections to the classical kink-like solutions and to the position of second-order phase transition with $\tau$ symmetry breaking  are found in the leading orders in gravitational constant, thin brane tension and in the critical point deviation scale.
\item The proper choice of gauge invariant variables of metric and scalar fluctuations around a non-trivial classical background has been elaborated and the gauge invariant action of second order in fluctuations has been obtained.
\item It is shown that an appropriate choice of metric and fluctuation variables  leads to the equations on scalar fluctuation spectrum suitable for perturbative calculations.
\item The nonperturbative discontinuous gravitational effects  in the mass spectrum of light localized scalar states has been studied in the presence of a thin-brane defect.
    When a defect is removed the absence of a massless Goldstone-like state is confirmed.
\item It is shown that in the coherent zero gravity and brane tension limit the branon field sector is characterized by an exactly solvable model.
\item In the presence of the thin brane with positive tension these nonperturbative effects are regularized and the spectrum contains a massless localized state and a heavy one. However the discontinuity still exists and it certainly deserves a more detailed non-perturbative investigation.
\item The thin brane with negative tension is the most curious case. The singular barriers form a potential well with infinitely tall walls. Thus we obtain the discrete spectrum of localized states confined to the well and completely isolated from the bulk. While that would be an ideal mechanism for matter localization on a brane the further study  is necessary  for its possible  generalizations to other fields.
\item In the broken $\tau$ symmetry phase a discontinuity in the  mass spectrum of the Higgs-like scalar fluctuation is discovered in the next-to-leading orders of critical point deviations and for vanishing defects.
\item As it was shown in \cite{partnuc} using $v \simeq 246 GeV$, the recently measured \cite{lhc},\cite{lhc1} Higgs mass $m \simeq 126GeV$ , $M_P \simeq 2.5\cdot 10^{18} GeV$ \cite{PDG} and the modern bound for the AdS curvature, $k > 0.004 eV$, \cite{adelb} one can obtain the following bounds for the scales and couplings of our model,
\ba
&&M > 3.5 TeV;\quad M_\ast > 3\cdot 10^8 GeV;\no
&& \kappa > 2\cdot 10^{- 15} .
\ea
Thus we conclude that the gravitational corrections on localization mechanism are indeed very small except for branon spectrum. \end{itemize}

\begin{acknowledgements}
We acknowledge the financial support by Grants RFBR, project 13-02-00127 and project 13-01-00136 as well as by the Saint Petersburg State University grant 11.38.660.2013. One of us (A.A.) was partially supported by projects FPA2010-20807.
\end{acknowledgements}


\begin{thebibliography}{99}
\bibitem{rushap} V.~A.~Rubakov and M.~E.~Shaposhnikov,
  V.~A.~Rubakov and M.~E.~Shaposhnikov,
  Phys.\ Lett.\ B {\bf 125} (1983) 136,
  Phys.\ Lett.\ B {\bf 125} (1983) 139.
\bibitem{otherbr} K.~Akama,
  Lect.\ Notes Phys.\  {\bf 176} (1982) 267
  [hep-th/0001113];\
\bibitem{otherbr1}
   M.~Visser,
  Phys.\ Lett.\ B {\bf 159} (1985) 22
  [hep-th/9910093];\
\bibitem{otherbr2}
   M.~Pavsic,
  Phys.\ Lett.\ A {\bf 116} (1986) 1
  [gr-qc/0101075];\
\bibitem{otherbr3}
   G.~W.~Gibbons and D.~L.~Wiltshire,
  Nucl.\ Phys.\ B {\bf 287} (1987) 717
  [hep-th/0109093].
\bibitem{ADD} N.~Arkani-Hamed, S.~Dimopoulos and G.~R.~Dvali,
  Phys.\ Lett.\ B {\bf 429} (1998) 263
  [hep-ph/9803315].
\bibitem{RSI}
  L.~Randall and R.~Sundrum,
  Phys.\ Rev.\ Lett.\  {\bf 83} (1999) 3370
  [hep-ph/9905221].
\bibitem{RSII} L.~Randall and R.~Sundrum,
  Phys.\ Rev.\ Lett.\  {\bf 83} (1999) 4690
  [hep-th/9906064].
\bibitem {RuBar} V.~A.~Rubakov,
  Phys.\ Usp.\  {\bf 44} (2001) 871
   [Usp.\ Fiz.\ Nauk {\bf 171} (2001) 913]
  [hep-ph/0104152];
\bibitem {RuBar1} V.~A.~Rubakov,
  Phys.\ Usp.\  {\bf 46} (2003) 211
   [Usp.\ Fiz.\ Nauk {\bf 173} (2003) 219].
\bibitem{rev1}A.~O.~Barvinsky,
  Phys.\ Usp.\  {\bf 48} (2005) 545
   [Usp.\ Fiz.\ Nauk {\bf 175} (2005) 569].
\bibitem{rev3}J.~L.~Hewett and M.~Spiropulu,
  Ann.\ Rev.\ Nucl.\ Part.\ Sci.\  {\bf 52} (2002) 397
  [hep-ph/0205106].
\bibitem{rev4}R.~Dick,
  Class.\ Quant.\ Grav.\  {\bf 18} (2001) R1
  [hep-th/0105320].
\bibitem{rev5}R.~Maartens,
  Living Rev.\ Rel.\  {\bf 7} (2004) 7
  [gr-qc/0312059].
\bibitem{rev6}P.~Brax, C.~van de Bruck and A.~-C.~Davis,
  Rept.\ Prog.\ Phys.\  {\bf 67} (2004) 2183
  [hep-th/0404011].
\bibitem{rev9} F.~Feruglio,
  Eur.\ Phys.\ J.\ C {\bf 33} (2004) S114
  [hep-ph/0401033].
\bibitem{rev11} C.~Csaki,
  In *Shifman, M. (ed.) et al.: From fields to strings, vol. 2* 967-1060
  [hep-ph/0404096].
\bibitem{loc6} V.~Dzhunushaliev, V.~Folomeev and M.~Minamitsuji,
  Rept.\ Prog.\ Phys.\  {\bf 73} (2010) 066901
  [arXiv:0904.1775 [gr-qc]].
\bibitem{rev121} O.~DeWolfe, D.~Z.~Freedman, S.~S.~Gubser and A.~Karch,
  Phys.\ Rev.\ D {\bf 62} (2000) 046008
  [hep-th/9909134].
\bibitem{rev13} M.~Gremm,
  Phys.\ Lett.\ B {\bf 478} (2000) 434
  [hep-th/9912060].
\bibitem{rev14} C.~Csaki, J.~Erlich, T.~J.~Hollowood and Y.~Shirman,
  Nucl.\ Phys.\ B {\bf 581} (2000) 309
  [hep-th/0001033].
\bibitem{rev15} M.~Giovannini,
  Phys.\ Rev.\ D {\bf 64} (2001) 064023
  [hep-th/0106041].
\bibitem{rev16} A.~Kehagias and K.~Tamvakis,
  Phys.\ Lett.\ B {\bf 504} (2001) 38
  [hep-th/0010112].
\bibitem{bron} K.~A.~Bronnikov and B.~E.~Meierovich,
  Grav.\ Cosmol.\  {\bf 9} (2003) 313
  [gr-qc/0402030].
\bibitem{aags1} A.~A.~Andrianov, V.~A.~Andrianov, P.~Giacconi and R.~Soldati,
  JHEP {\bf 0307} (2003) 063
  [hep-ph/0305271].
\bibitem{rev17}  D.~Bazeia and A.~R.~Gomes,
  JHEP {\bf 0405} (2004) 012
  [hep-th/0403141];
\bibitem{rev17_1}  D.~Bazeia, C.~Furtado and A.~R.~Gomes,
  JCAP {\bf 0402} (2004) 002
  [hep-th/0308034].
\bibitem{Bazeia:2013euc}
  D.~Bazeia, A.~S.~Lobao, L.~Losano and R.~Menezes,
  Phys.\ Rev.\ D {\bf 88}, 045001 (2013)
  [arXiv:1306.2618 [hep-th]].
\bibitem{rev181}  A.~de Souza Dutra and A.~C.~Amaro de Faria, Jr.,
  Phys.\ Lett.\ B {\bf 642} (2006) 274
  [hep-th/0610315].
\bibitem{rev12} S.~L.~Dubovsky, V.~A.~Rubakov and P.~G.~Tinyakov,
  Phys.\ Rev.\ D {\bf 62} (2000) 105011
  [hep-th/0006046].
\bibitem{rev18} M.~Shaposhnikov, P.~Tinyakov and K.~Zuleta,
  Phys.\ Rev.\ D {\bf 70} (2004) 104019
  [hep-th/0411031].
\bibitem{aags2} A.~A.~Andrianov, V.~A.~Andrianov, P.~Giacconi and R.~Soldati,
  JHEP {\bf 0507} (2005) 003
  [hep-th/0503115].
\bibitem{Ahmed:2012nh}
  A.~Ahmed and B.~Grzadkowski,
  JHEP {\bf 1301}, 177 (2013)
  [arXiv:1210.6708 [hep-th]].
\bibitem{rev19} A.~A.~Andrianov and L.~Vecchi,
  Phys.\ Rev.\ D {\bf 77} (2008) 044035
  [arXiv:0711.1955 [hep-th]].
\bibitem{partnuc} A.~A.~Andrianov, V.~A.~Andrianov and O.~O.~Novikov,
  Phys.\ Part.\ Nucl.\  {\bf 44}, 190 (2013)
  [arXiv:1210.3698 [hep-th]].
\bibitem{branon} J.~A.~R.~Cembranos, A.~Dobado and A.~L.~Maroto,
  Phys.\ Rev.\ D {\bf 70} (2004) 096001
  [hep-ph/0405286].

\bibitem{loc1} R.~Koley and S.~Kar,
  Class.\ Quant.\ Grav.\  {\bf 22} (2005) 753
  [hep-th/0407158].
\bibitem{loc11} A.~Melfo, N.~Pantoja and J.~D.~Tempo,
  Phys.\ Rev.\ D {\bf 73} (2006) 044033
  [hep-th/0601161];
\bibitem{loc11_1} O.~Castillo-Felisola and I.~Schmidt,
  Phys.\ Rev.\ D {\bf 82}, 124062 (2010)
  [arXiv:1008.1281 [hep-th]].
\bibitem{loc12} D.~Bazeia, F.~A.~Brito and R.~C.~Fonseca,
  Eur.\ Phys.\ J.\ C {\bf 63} (2009) 163
  [arXiv:0809.3048 [hep-th]].
\bibitem{loc2} Y.~-X.~Liu, H.~-T.~Li, Z.~-H.~Zhao, J.~-X.~Li and J.~-R.~Ren,
  JHEP {\bf 0910} (2009) 091
  [arXiv:0909.2312 [hep-th]].
\bibitem{loc22}  Z.~-H.~Zhao, Y.~-X.~Liu, H.~-T.~Li and Y.~-Q.~Wang,
  Phys.\ Rev.\ D {\bf 82} (2010) 084030
  [arXiv:1004.2181 [hep-th]].
\bibitem{loc23} H.~-T.~Li, Y.~-X.~Liu, Z.~-H.~Zhao and H.~Guo,
  Phys.\ Rev.\ D {\bf 83} (2011) 045006
  [arXiv:1006.4240 [hep-th]].
\bibitem{loc3} C.~A.~S.~Almeida, M.~M.~Ferreira, Jr., A.~R.~Gomes and R.~Casana,
  Phys.\ Rev.\ D {\bf 79} (2009) 125022
  [arXiv:0901.3543 [hep-th]].
\bibitem{loc4} C.~-EFu, Y.~-X.~Liu and H.~Guo,
  Phys.\ Rev.\ D {\bf 84} (2011) 044036
  [arXiv:1101.0336 [hep-th]].
\bibitem{loc44} M.~N.~Smolyakov,
  Phys.\ Rev.\ D {\bf 85} (2012) 045036
  [arXiv:1111.1366 [hep-th]].
\bibitem{Kirpichnikov:2013cza}
  D.~V.~Kirpichnikov,
  arXiv:1310.2866 [hep-th].
\bibitem{loc5} L.~B.~Castro,
  Phys.\ Rev.\ D {\bf 83} (2011) 045002
  [arXiv:1008.3665 [hep-th]].

\bibitem{GW} G.~W.~Gibbons, S.~W.~Hawking,
  Phys.\ Rev.\ D {\bf 15} (1977) 2752.
\bibitem{york} J.~W.~York,
  Phys.\ Rev.\ D {\bf 28} (1972) 1082.
\bibitem{Lewin} L.~Lewin,
  \emph{Polylogarithms and Associated Functions},
  North Holland, New York, Oxford (1981), pg. 1.
\bibitem{bar} J.M.~Bardeen,
Phys.\ Rev.\ D {\bf 22}, 1882 (1980).
\bi{susy1} F.~Cooper, A.~Khare and U.~Sukhatme,
  Phys.\ Rept.\  {\bf 251} (1995) 267
  [hep-th/9405029].
\bi{susy2} A.~A.~Andrianov and M.~V.~Ioffe,
  J.\ Phys.\ A {\bf 45}, 503001 (2012)
  [arXiv:1207.6799 [hep-th]].

\bibitem{lhc}
  S.~Chatrchyan {\it et al.}  [CMS Collaboration],
  Phys.\ Lett.\ B {\bf 716} (2012) 30
  [arXiv:1207.7235 [hep-ex]].
\bibitem{lhc1} G.~Aad {\it et al.}  [ATLAS Collaboration],
  Phys.\ Lett.\ B {\bf 716} (2012) 1
  [arXiv:1207.7214 [hep-ex]].
\bibitem{PDG}
  J.~Beringer {\it et al.}  [Particle Data Group Collaboration],
  Phys.\ Rev.\ D {\bf 86} (2012) 010001.
\bibitem{adelb} E.~G.~Adelberger, J.~H.~Gundlach, B.~R.~Heckel, S.~Hoedl and S.~Schlamminger,
  Prog.\ Part.\ Nucl.\ Phys.\  {\bf 62} (2009) 102.
\end{thebibliography}
\end{document}